\documentclass[aps,pre,twocolumn,nofootinbib,showpacs,preprintnumbers,amsmath,amssymb,
superscriptaddress,floatfix,showkeys]{revtex4-1}

\usepackage{graphicx}
\usepackage{dcolumn}
\usepackage{bm}
\usepackage{color} 
\usepackage{amsfonts}
\usepackage{amsmath}
\usepackage{amssymb}
\usepackage{amsthm}
\usepackage{hyperref}
\usepackage[dvipsnames]{xcolor}
\usepackage{soul} 

\newcommand{\unnec}[1]{}

\newcommand{\unnecc}[1]{}

\begin{document} 
 
 
\title{Phase Boundaries of Bulk 2D Rhombi$^{\dagger}$}

\author{Gerardo Odriozola}
\email[]{godriozo@azc.uam.mx}
\affiliation{\'Area de F\'isica de Procesos
Irreversibles, Divisi\'on de Ciencias B\'asicas e Ingenier\'ia, Universidad
Aut\'onoma Metropolitana-Azcapotzalco, Av. San Pablo 180, 02200 CD M\'exico,
Mexico}

\author{P\'eter Gurin} 
\affiliation{Physics Departement, Centre for Natural Sciences, University of Pannonia, 
P.O. Box 158, Veszpr\'em H-8201, Hungary}


\begin{abstract}
We conducted replica exchange Monte Carlo simulations to investigate the phase diagram of identical hard rhombi systems in two-dimensions. The rhombi’s shape varies from nearly square-like, as their minor angle $a$ approaches $90^\circ$, to needle-like, as it approaches $0^\circ$. For angles near $90^\circ$, we get an isotropic fluid, a rhombatic fluid, a rotator phase, and a columnar space-filling structure with increasing density. Conversely, as $a$ approaches $0^\circ$, the results resemble the needle limit. Even for angles as small as $a=$ $20^\circ$, we still obtain isotropic, nematic, and rhombatic fluids before reaching a rhombic solid, but the nematic phase gains importance with decreasing $a$. At $a \approx $ $60^\circ$, aperiodic space-filling structures with long-range six-fold orientational symmetry dominate over periodic candidates such as the rhombic and rhombille. This aperiodic solid undergoes a melting process leading to a phase with quasi-long-range six-fold orientational symmetry, a hexatic fluid, before reaching the isotropic phase.
\end{abstract} 


\maketitle

\onecolumngrid

\begin{center}
\footnotesize
\textit{
Accepted for publication in Computational Materials Science.
This is a preprint of an article published in Computational Materials Science.
DOI: \href{https://doi.org/10.1016/j.commatsci.2024.112919}{10.1016/j.commatsci.2024.112919}
}
\end{center}

\twocolumngrid

\vspace{0.3cm}

\section{Introduction}

\begin{figure}[t!]
\centering
\includegraphics[width=0.95\linewidth]{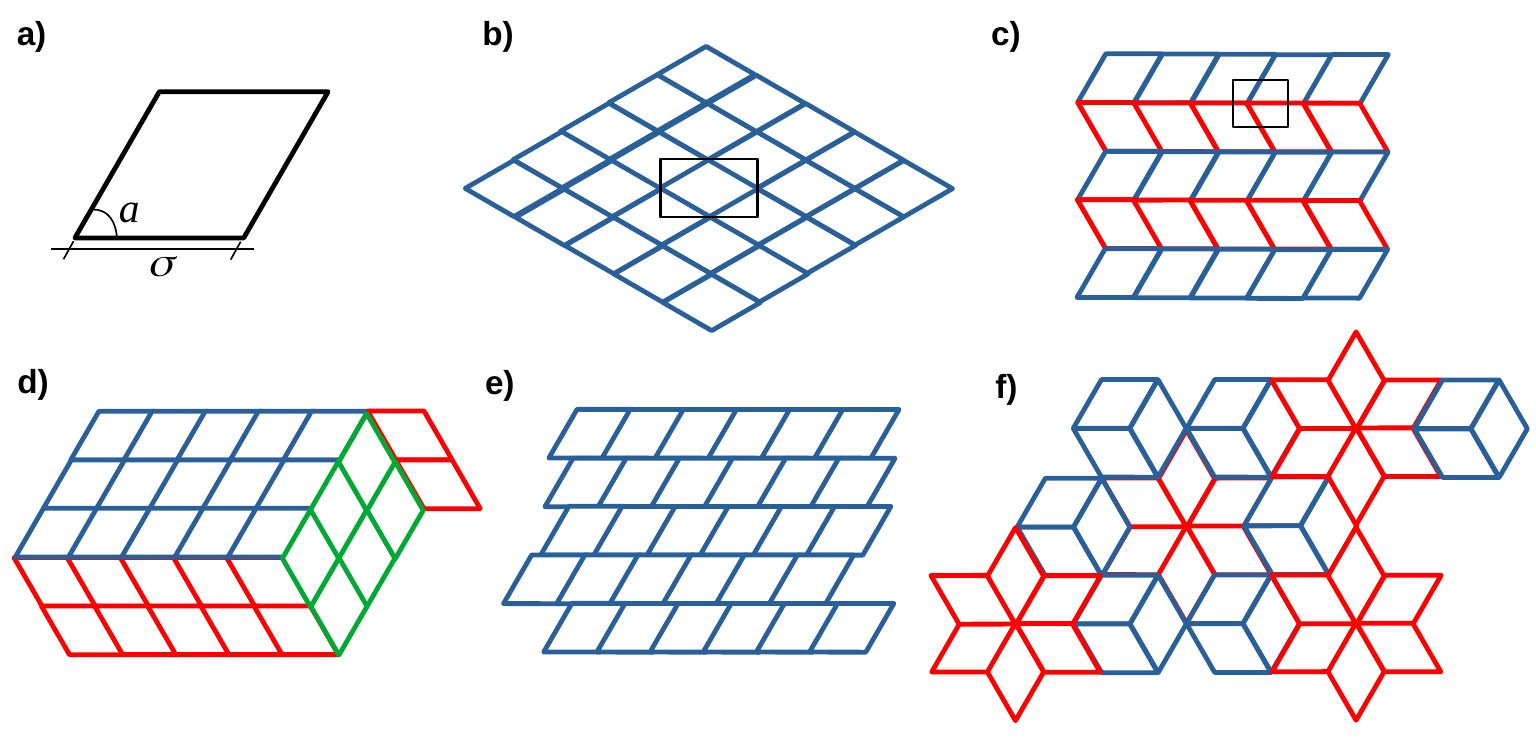}
\caption{ a) Rhombus with side $\sigma$ and minor angle $a$. As $\sigma$ is taken as the unit of length, the only relevant parameter defining the hard rhombi system is $a$. For all panels in this figure we have set $a=60^\circ$. b) Periodic rhombic tiling (also known as centered rectangular tiling). The rectangular unit cell is outlined with black lines. c) Another possible periodic rectangular tiling. d) Poly-rhombic pattern (locally rhombic but with several different rhombic-like regions). e) Columnar arrangement. f) Rhombille~\cite{conway2008} (also called tumbling blocks~\cite{smith2002}) periodic pattern. In panels c) and d), rhombi are colored according to their orientation, whereas in panel f), they are colored according to their local pattern (blue for cube-like shapes and red for stars).
}
\label{fig:General}
\end{figure}

As rectangles, rhombi serve as a unique generalization, incorporating features of both squares and needles.
This versatility enables them to convolutedly populate two-dimensional Euclidean space with an array of patterns, spanning from regular periodic arrangements to more intricate aperiodic configurations~\cite{grunbaum1987tilings,Whitelam2012,Whitelam2015}. Unsurprisingly, rhombille tiling patterns, with their distinctive star and cubic-like shapes, have found applications in art and design spanning from ancient Greece, as seen in the floors at Delos~\cite{dunbabin1999mosaics}, to the modern day, including the cruise terminal in Tallinn  (see panel f) of Fig.~\ref{fig:General}). They have also left their mark in medieval times, as evidenced by their presence in the Siena Cathedral~\cite{wallis1902italian}. These captivating patterns arise particularly when rhombi possess a minor angle $a=60^\circ$~\cite{Whitelam2012,Stannard2012,Whitelam2015} (see Fig.~\ref{fig:General} a) for a scheme showing the definitions of the minor angle $a$ and the side length $\sigma$). However, when $a$ differs from the special values of $90^\circ$ and $60^\circ$, an intriguing interplay emerges. These diverse rhombi have the capacity to carve out their spatial arrangements through various means, including adopting either a periodic rhombic tiling  (see panels b) and c) of Fig.~\ref{fig:General}), a poly-rhombic configuration (Fig.~\ref{fig:General} d)), or organizing into distinct columnar structures (panel e) of Fig.~\ref{fig:General}).

When the angle $a$ assumes a value of $90^\circ$, the rhombi perfectly transform into squares, and their packing results in a square crystalline phase~\cite{Wojciechowski04,Walsh16,Anderson2017}. This arrangement is marked by a striking symmetry, where both columns and rows are alike, defining mutually orthogonal directions. Although hard squares could potentially adopt a columnar arrangement, the tendency for rows to be lost due to relative columnar motion is counterbalanced by the prevailing dominance of the square tiling. With decreasing pressure, the behavior of hard squares takes on an even more captivating dimension. It unfolds as a two-step melting process of the KTHNY type~\cite{berezinskii1971,berezinskii1972,Kosterlitz-Thouless,Halperin-Nelson,Anderson2017}, characterized by consecutive high-order transitions. This intricate dance of phases entails the solid and isotropic fluid enveloping an intermediate $i$-atic phase, the tetratic.

Note that the realm of rhombic tilings is not exclusive to hard rhombi. Recent studies have unveiled that hard rounded squares~\cite{Zhao2011,Avendano12,Zhaglin_ChinPhysB_2018} and superdisks~\cite{Jiao08,Rossi5286,Meijer-et.al_NAT.COMM_2017}, too, possess the ability to spontaneously arrange themselves, giving rise to two distinct rhombic tilings. In the scenario where superdisks closely resemble squares, the rhombic tiling gradually transforms into a square arrangement. This evolution induces a subtle distortion in the intermediate tetratic~\cite{Anderson2017,Donev2006,martinez2009enhanced}, leading to a fluid with quasi-long-range particle orientation and bond orientational orders which inherits the properties of the rhombic primitive unit cell (rhombatic)~\cite{gurin2020three}. Conversely, as superdisks take on a disk-like nature, the rhombatic fluid undergoes a metamorphosis into a hexatic fluid, eventually transitioning into an isotropic fluid with decreasing system density. Remarkably, at the midway point between squares and disks, a curious melting phenomenon unfolds, progressing from rhombic to rhombatic, then from rhombatic to hexatic, culminating in the emergence of the isotropic fluid~\cite{gurin2020three}.

As $a$ approaches zero, rhombi transform into needle-like shapes. Consequently, a needle-like behavior is anticipated for small $a$ values, implying that particles are likely to orient themselves forming a nematic phase at relatively low densities~\cite{Onsager_AnnNYAcadSci_1949,kayser1978,Frenkel-Eppenga_PRA_1985,tarjus1991new,Vink2009,Gurin_PRE_2011,foulaadvand2011}. Moreover, the range of packing fractions at which the nematic phase prevails is anticipated to expand as $a$ decreases. Conversely, when $a$ approaches $90^\circ$, a return to the hard-square limit is expected, akin to the behavior exhibited by superdisks. This suggests the resurgence of a square solid and a tetratic fluid under these conditions. Adding a layer of complexity, the case $a=60^\circ$ presents a unique scenario. In this case, predicting the thermodynamically stable arrangement at high densities proves challenging, given the interplay between non-periodic space-filling configurations and their periodic counterparts. This competition, driven by entropy and characteristic of hard models, defies easy prediction regarding the prevailing configuration. Additionally, the solid candidate that emerges victorious might undergo a melting process, introducing dislocations followed by disclinations, a phenomenon common to hard convex particles~\cite{Wojciechowski04,Donev2006,MartinezRaton2006,Avendano12,Kapfer2015,Anderson2017,gurin2020three,martinez2020orientational,hou2020emergent,velasco2023prediction} (though hard concave particles may exhibit different behavior~\cite{avendano2017packing,ramirez2023densest}), eventually leading to the emergence of the corresponding $i$-atic fluid.

The focus of this study lies in constructing the phase diagram of hard rhombi. To achieve this objective, we have engaged in a series of replica exchange Monte Carlo (REMC) simulations, involving relatively modest system sizes comprising $N=500$ particles. Our research encompasses the computation of equations of state (EOS) and the isothermal compressibility across various $a$ values, alongside the analysis of bond-order and global order parameters. Moreover, our exploration entails a survey through several snapshots and their corresponding static structure factors (SSFs). In some instances where deeper insights were needed, simulations involving $N=5000$ particles were conducted to illuminate the decay of positional and bond-orientational correlations. As anticipated, the resultant phase diagram unfurls a rich complexity, revealing a diverse array of fluid and solid phases. Our results unveil that transitions like isotropic-nematic, isotropic-rhombatic, isotropic-hexatic, and nematic-rhombatic are characterized by discernible peaks in isothermal compressibility across all cases. This distinctive trait aids in readily locating these phase diagram boundaries. Conversely, the task of identifying fluid-solid boundaries proves more intricate due to the existence of columnar and aperiodic solid phases, with the lack of long or quasi-long-range correlations, and the absence of conspicuous isothermal compressibility peaks.

\section{Simulation details}

We implemented the replica exchange Monte Carlo (REMC) technique~\cite{Marinari92, Lyubartsev92, hukushima96} to minimize hysteresis as much as possible. This technique is based on an extended ensemble with partition function $Q_{\text{ext}} = \prod_{i=1}^{n_r} Q_{i}$, where $Q_i$ represents the partition function of ensemble $i$, and $n_r$ is the number of ensembles, which corresponds to the number of sampling replicas. The use of $Q_{\text{ext}}$ enables the execution of swap trial moves between any two replicas while satisfying the detailed balance condition.

Given that we are working with hard particles, it is advantageous to expand the isobaric-isothermal ensembles with respect to pressure~\cite{odriozola2009replica}. Consequently, the partition function of the extended ensemble can be expressed as~\cite{Okabe01, odriozola2009replica}:
\begin{equation}
Q_{\text{ext}} = \prod_{i=1}^{n_r} Q_{NTP_i}.
\end{equation}
Here, $Q_{NTP_i}$ signifies the partition function of the isobaric-isothermal ensemble for a system containing $N$ particles at temperature $T$ and pressure $P_i$ (force per unit length in the 2D case).

The $NTP_i$ ensembles can be sampled through a standard Monte Carlo (MC) implementation, involving independent trial displacements, rotations, volume changes, and distortions of the system cells. For this purpose, we have implemented a segment-segment check to detect overlaps. Also, we have employed event-chain MC moves~\cite{bernard2009event, michel2014generalized} instead of conventional random displacements to improve performance at high densities. For hard particles, these moves allow us to accept all trials while maintaining the detailed balance condition~\cite{bernard2009event}. However, our rotation trials remain conventional. Furthermore, the sampling includes trial changes in the angles and relative lengths of the cell lattice vectors, which increases the degrees of freedom for our relatively small systems ($N \sim 500$).

Swap trials are conducted by considering homogeneously distributed probabilities for selecting any adjacent pairs of replicas. To determine the acceptance of a swap trial, we apply the following rule~\cite{Okabe01, odriozola2009replica}:
\begin{equation}
\label{accP}
P_{\text{acc}} = \min\left(1, \exp\left[\beta(P_i - P_j)(A_i - A_j)\right]\right).
\end{equation}
Here, $A_i - A_j$ represents the area difference between replicas $i$ and $j$, and $\beta = 1 / (k_B T)$ is the reciprocal temperature. Adjacent pressures must be sufficiently close to ensure reasonable swap acceptance rates.

We investigate systems consisting of $N$ identical rhombi with a side length of $\sigma$, a minor angle denoted as $a$, and an area given by $\sigma^2 \sin{(a)}$. The angle $a$ is systematically varied from $20^\circ$ to $90^\circ$ in increments of $5^\circ$. For our analysis, we fix $N = 500$ for quantities such as $Z = \beta P /\rho$, where $\rho = N/A$ represents the number density ($\eta=\rho \sigma^2 \sin(a)$ is the packing fraction).  We also compute $\chi = N(\langle\rho^2\rangle - \langle\rho\rangle^2)/\langle\rho\rangle^2$, and the global orientational order parameter $P_n = (\langle 1/N \sum_i^N \cos(n\theta_i) \rangle^2 + \langle 1/N \sum_i^N \sin(n\theta_i) \rangle^2)$, where $\theta_i$ represents the angle of particle $i$ in relation to an arbitrary fixed frame. The values of $n$ used are 2, 4, and 6 to detect two, four, and six-fold orientational symmetry. In the context of global bond-orientational order parameters, we define $\Psi_n = \langle 1/N |\sum_i^N \varphi_{n,i}| \rangle$, with $n = 4$, 6, 8, 10, 12, and $\varphi_{n,i} = 1/N_i^b \sum_j^{N_i^b} \exp{(n\theta_{ij}\sqrt{-1})}$. In this expression, $N_i^b$ represents the number of neighbors of particle $i$ determined by a cutoff such that $\langle N_i^b \rangle$ converges to $n$.

When evaluating the decay of positional and bond-orientational correlations with distance, we use $N = 5000$. Positional correlations are constructed from the peaks of the radial distribution function based on distance. On the other hand, bond-orientational correlations are computed by assigning the complex number $\varphi_{n,i}$ to each particle $i$ and subsequently utilizing it in the formula: $g_n(r) = \langle\sum_{i\neq j}\delta(r-r_{ij})\varphi_{n,i}\varphi_{n,j}^*/\sum_{i\neq j}\delta(r-r_{ij})\rangle$.

\section{Results}

\begin{figure}[t!]
\centering
\includegraphics[width=0.9\linewidth]{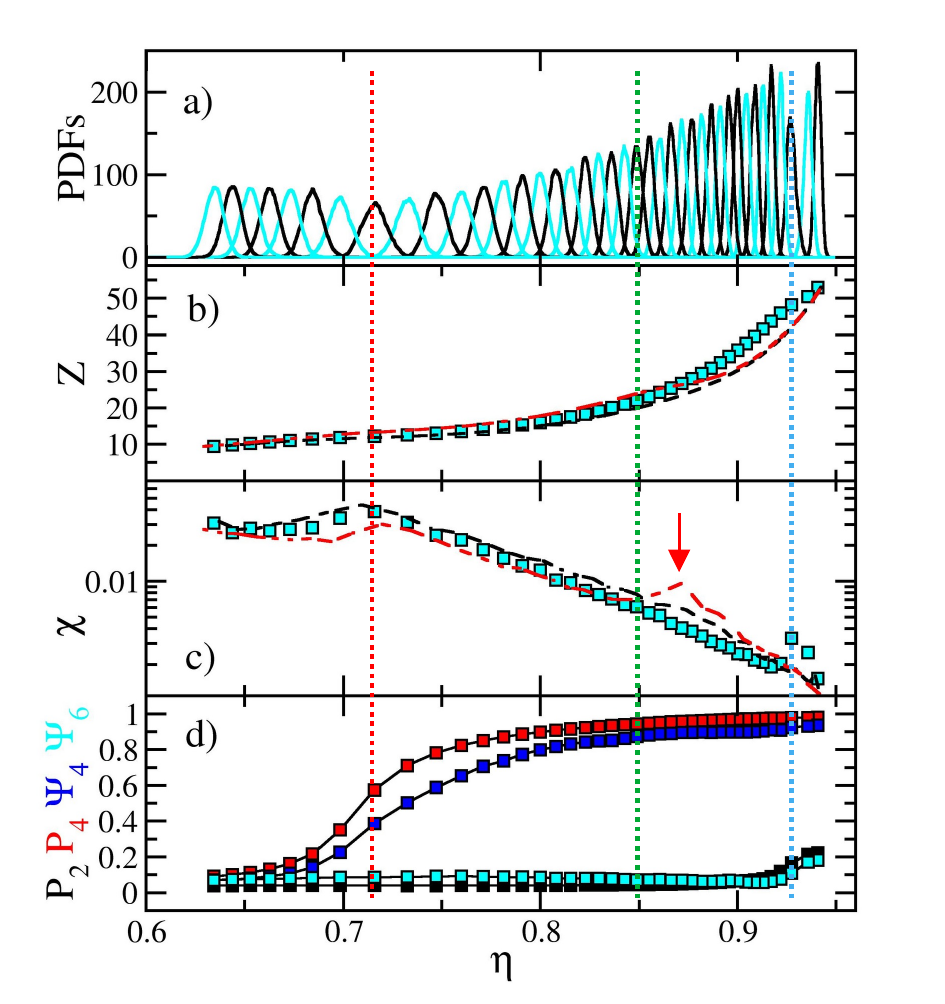}
\caption{a) Probability density functions, b) compressibility factor, c) dimensionless isothermal compressibility, and d) order parameters as a function of the packing fraction for rhombi with a minor angle $a=85^\circ$. Black and red dashed lines in panels b) and c) correspond to angles $a=90^\circ$ and $80^\circ$, respectively. In panel d), black and red symbols depict orientational order parameters $P_2$ and $P_4$, respectively, while blue and cyan symbols correspond to bond-orientational order parameters $\Psi_4$ and $\Psi_6$, respectively. The vertical dotted lines signal phase transitions for the $a=85^\circ$ case. The red and cyan lines denote isotropic-rhombatic fluid-fluid and rotator-columnar solid-solid transitions, respectively. The green vertical line corresponds to the fluid-solid transition obtained from the building of quasi-long-range positional correlations and long-range bond-orientational correlations.
}
\label{fig:EOS-a85}
\end{figure}

We commence our analysis with $a = 85^\circ$, as we anticipate that these rhombi would yield results similar to those of hard squares~\cite{Walsh16,Anderson2017}. The probability density functions (PDFs), as well as the compressibility factor $Z$, dimensionless isothermal compressibility $\chi$, and the order parameters $P_2$, $P_4$, along with the bond-order parameters $\Psi_4$ and $\Psi_6$, are illustrated in Fig.~\ref{fig:EOS-a85}. In panels b) and c), we compare the acquired data with the case of $a = 90^\circ$, depicted as black dashed lines. Upon comparison, we observe marginal distinctions for packing fractions $\eta<0.75$, except a slight shift in the $\chi$ peak at low density. This shift is further evidenced by the red dashed curves corresponding to $a = 80^\circ$, although in this instance, the shift is more pronounced.

\begin{figure}[t!]
\centering
\includegraphics[width=0.9\linewidth]{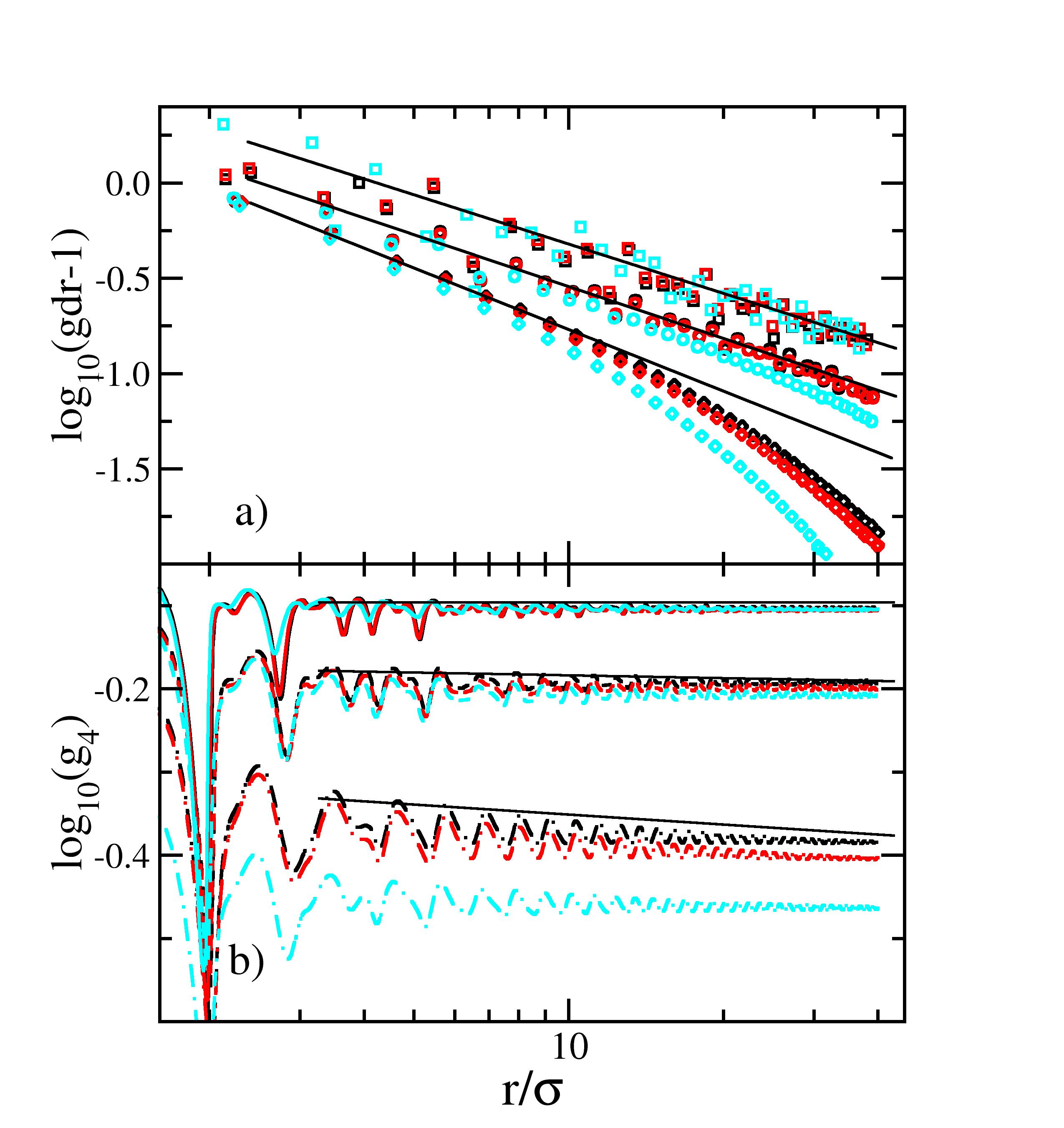}
\caption{a) $\log_{10}(gdr-1)$, with $gdr$ being the peaks of the radial distribution function, and b) $\log_{10}(g_4)$ as a function of distance. Both panels correspond to systems with $N=5000$, and for $a= 90^{\circ}$ (black), $85^{\circ}$ (red), and $80^{\circ}$ (cyan). Squares in panel a) and solid lines in b) correspond to $\eta=0.848$, 0.850, and 0.872 for decreasing $a$. Circles and short dashed lines correspond to $\eta=0.800$, and diamonds and dash-dotted lines correspond to $\eta=0.760$. The straight lines in the plots serve as visual guides. The upper line in panel b) is horizontal. }
\label{fig:correlations-a80-90}
\end{figure}

The $\chi$ peak observed around $\eta \approx 0.71$ corresponds with an increase in $P_4$ and $\Psi_4$, indicative of the high-order isotropic-rhombatic (tetratic for $a = 90^\circ$) fluid-fluid transition~\cite{Walsh16,Anderson2017}. These $\chi$ peaks result from the broadening of the PDFs as shown in panel a) of Fig.~\ref{fig:EOS-a85}. Note that the peaks of the PDFs keep their Gaussian shape, suggesting high-order transitions in line with Anderson et al.~findings~\cite{Anderson2017}. The position of this $\chi$ peak is marked by the vertical red dotted line in Fig.~\ref{fig:EOS-a85}. Similar to the behavior observed in hard squares, the $i$-atic phase extends over a relatively wide $\eta$ range up to $\eta \approx 0.85$, supported by the decay of positional correlations (and four-fold bond-orientational correlations) shown in Fig.~\ref{fig:correlations-a80-90}. Thus, the rhombatic phase lies between the vertical red and green lines delineated in Fig.~\ref{fig:EOS-a85}.

\begin{figure*}[t!]
\centering
\includegraphics[width=0.9\linewidth]{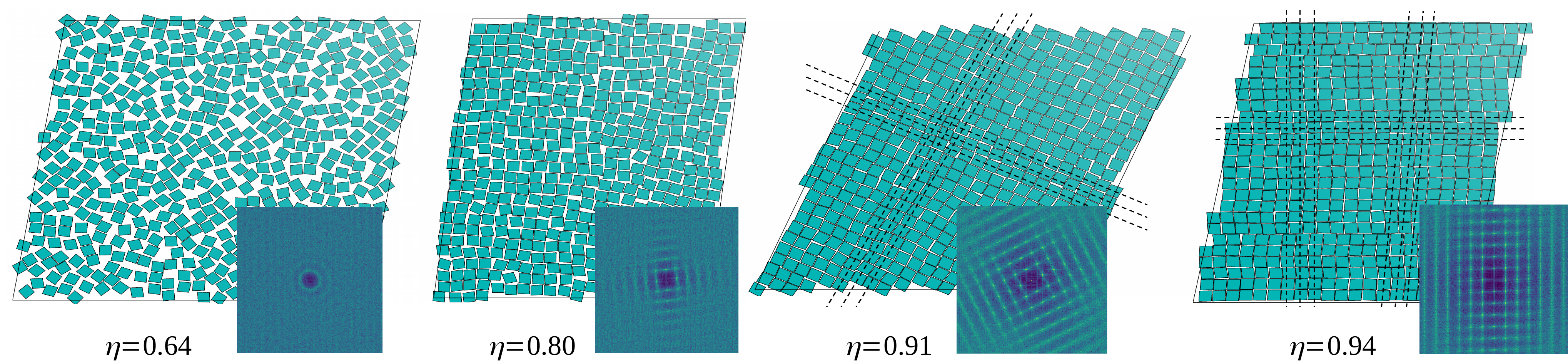}
\caption{Snapshots and their corresponding SSFs (computed by using freud~\cite{ramasubramani2020freud}) for rhombi with $a=85^\circ$ at increasing packing fraction. In the third snapshot, dashed lines form angles of $85^\circ$, following the columns and rows formed by the particles. In the fourth snapshot, dashed lines form angles of $90^\circ$ (left) and $85^\circ$ (right), with only the horizontal lines aligning along the columns (horizontal in this case) formed by the particles.  }
\label{fig:snapshots-a85}
\end{figure*}

At very large $\eta$, we encounter another $\chi$ peak occurring around $\eta \approx 0.93$, accompanied by a small yet distinct increase in $P_2$ and $\Psi_6$. This peak is absent in the case of squares, indicating a qualitatively different behavior in rhombi induced by a slight alteration of $a$, which breaks the tetratic symmetry exhibited by squares. The vertical cyan dashed line in Fig.~\ref{fig:EOS-a85} highlights this peak.
The rise in $P_2$ implies that the solid transforms from a lattice where particles can either point left-right or up-down without distinction, preserving four-fold and destroying two-fold symmetry, to another lattice where a stronger two-fold correlation is present. We term this low-density solid phase as 'rotator' phase, although it's important to note that it differs from the one observed for ellipses~\cite{Cuesta1990,Donev05b,Bautista14} with an aspect ratio close to one. In such cases, particles can rotate with minimal hindrance from their neighbors. In contrast, changing the orientation of particles in the rhombic rotator structure is notably more difficult, as only two directions are entirely allowed (90-degree turns). The high-density $\chi$ peak is also observed in the case of $a=80^\circ$, although it emerges at significantly lower densities, around $\eta \approx 0.87$. This peak is indicated by a red arrow and corresponds to the red dashed curve. Furthermore, the peak aligns with a plateau in $Z$, as depicted by the red dashed curve in panel b), before approaching the black dashed line corresponding to $a=90^\circ$.

2D systems of short-range interacting particles at finite temperature and pressure always show a decay of positional correlations~\cite{peierls1934uber,mermin1966,berezinskii1971,berezinskii1972}. Panel a) of Fig.~\ref{fig:correlations-a80-90} illustrates the logarithmic decay of peaks in the radial distribution function. The results are presented for three cases: $a=90^\circ$ (black symbols), $85^\circ$ (red symbols), and $80^\circ$ (cyan symbols), employing $N=5000$. For $\eta$ values of 0.76 and 0.80, we use diamond and circle markers, respectively, while we employ squares to illustrate the results for $\eta=$ 0.848, 0.850, and 0.872 with decreasing $a$. Correlations exhibit a swifter decay as $a$ decreases for a fixed $\eta$. This observation leads us to anticipate an increase in the $\eta$ value, marking the fluid-solid transition as $a$ diminishes. Through large-scale simulations of hard squares, Anderson et al.~\cite{Anderson2017} identified the tetratic-solid transition at $\eta=0.848$, which corresponds to the lowest $\eta$ value displaying an algebraic decay in the radial distribution function peaks. As shown in Fig.~\ref{fig:correlations-a80-90} a), distinguishing this type of decay from the curves corresponding to $\eta=0.80$ proves challenging. Therefore, to determine the $\eta$ of the fluid-solid transition for $85^\circ$ and $80^\circ$, we simply sought the $\eta$ values that exhibited similar trends to the one corresponding to $a=90^\circ$. These values turned out to be 0.850 and 0.872 for $85^\circ$ and $80^\circ$, respectively.

An alternative approach to distinguish between a solid and a fluid phase relies on the long-range behavior of bond-orientational correlations. Solids exhibit long-range correlations, while $i$-atic fluids display quasi-long-range correlations. We present a log-log plot of these four-fold correlations for the same $a$ and $\eta$ values as shown in Fig.~\ref{fig:correlations-a80-90} a) in panel b). While retaining the same color code for identical $a$ values, we have replaced the square, circle, and diamond symbols with solid, dashed, and dotted-dashed lines, respectively. Consequently, horizontal curves are anticipated only for the highest densities, whereas the other curves should exhibit negative slopes. Once again, distinguishing a horizontal trend for the curves corresponding to $\eta=0.8$ remains challenging. However, it is notably simpler to align the curves for $a=90^\circ$ with $\eta=0.848$. Nonetheless, there appears to be a slight negative slope for $\eta=0.8$, which increases its absolute value for $\eta=0.76$. We have incorporated horizontal and negatively sloped lines as visual aids for each group of curves.

For the $a=85^\circ$ case, we present snapshots and their corresponding SSFs, as obtained from freud software~\cite{ramasubramani2020freud}, with increasing density from left to right in Fig.~\ref{fig:snapshots-a85}. These snapshots and SSFs reveal an isotropic fluid in the leftmost panel, followed by a rhombatic fluid, and then two distinct high-density phases appearing at $\eta=0.91$ and $0.94$. In the third panel, the particles define columns and rows at an angle of $85^\circ$, aligning with $a$. This snapshot depicts particles primarily pointing indistinctly in two orthogonal directions, preserving four-fold symmetry while abolishing the two-fold symmetry. In contrast, at the highest density, particles predominantly point in a single direction (either left or right in this context), forming columns. Thus, within a given column, most particles align in the same direction, enhancing packing efficiency. It's important to observe that we cannot define rows in this snapshot (dashed lines forming $90^\circ$ and $85^\circ$ angles with the column direction have been included as visual guides), as columns can shift horizontally without strongly interacting with their neighbors. This configuration doesn't maintain quasi-long-range positional correlations or long-range bond-orientational correlations.

\begin{figure}[t!]
\centering
\includegraphics[width=0.9\linewidth]{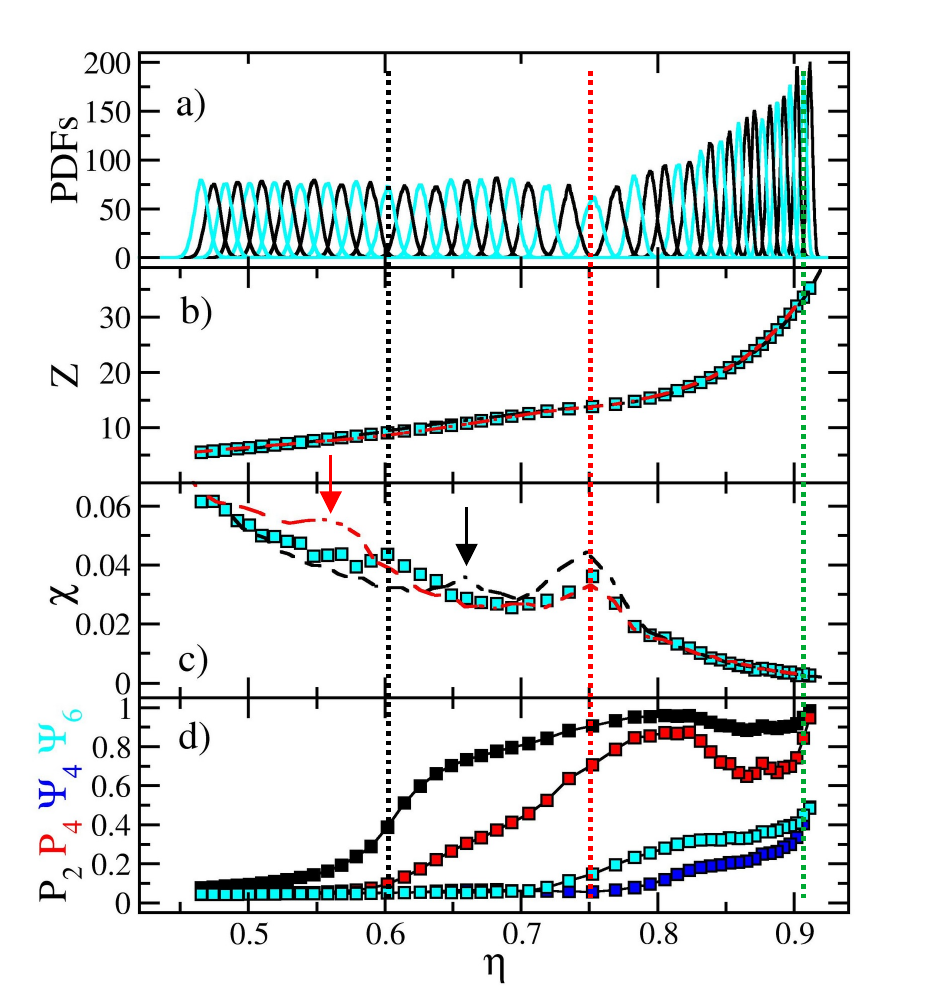}
\caption{a) Probability density functions, b) compressibility factor, c) dimensionless isothermal compressibility, and d) order parameters as a function of the packing fraction for rhombi with a minor angle $a=40^\circ$. Black and red dashed lines in panels b) and c) correspond to angles $a=45^\circ$ and $35^\circ$, respectively. In panel d), black and red symbols depict orientational order parameters $P_2$ and $P_4$, respectively, while blue and cyan symbols correspond to bond-orientational order parameters $\Psi_4$ and $\Psi_6$, respectively. The vertical dotted lines signal phase transitions for the $a=40^\circ$ case. The black and red lines denote isotropic-nematic and nematic-rhombatic fluid-fluid transitions, respectively. The green vertical line corresponds to the fluid-solid transition obtained from the melting of a rhombic solid. The black and red arrows point out $\chi$ peaks for the $a=45^\circ$ and $35^\circ$ curves, respectively.   }
\label{fig:EOS-a40}
\end{figure}

At low $a$ values, we anticipate the emergence of a low-density nematic fluid. As $a$ increases from low values, the nematic phase should manifest at higher packing fractions, with a subsequent point of disappearance at sufficiently large $a$. Let's examine the behavior around $a=40^\circ$. Fig.~\ref{fig:EOS-a40} presents four panels showcasing the PDFs, $Z$, $\chi$, and the global order parameters $P_2$, $P_4$, $\Psi_4$, and $\Psi_6$ for $a=40^\circ$. Similar to Fig.~\ref{fig:EOS-a85}, we have included the cases for $a=45^\circ$ (represented by black dashed lines) and $a=35^\circ$ (represented by red dashed lines) in panels b) and c) to observe any shifts in the $\chi$ peaks, if present.

The PDFs exhibit two distributions that are broader and shorter than their neighboring distributions while maintaining a Gaussian shape. This suggests the occurrence of higher-order transitions. These distributions correspondingly lead to a reduction in the slope of $Z$ and peaks in $\chi$. A relatively small $\chi$ peak at $\eta \approx 0.6$ coincides with a sharp increase in $P_2$, signaling the emergence of the nematic phase. The second peak arises around $\eta \approx 0.75$, accompanied by heightened values of $P_4$ and $\Psi_6$, marking the development of the rhombatic phase. It's worth noting that, in general, both $\Psi_4$ and $\Psi_6$ increase during the formation of the rhombatic fluid, with their increments dependent on the value of $a$. As $a$ increases, $\Psi_4$ is favored over $\Psi_6$, whereas the opposite occurs when $a$ decreases. We mark these two transitions with vertical black and red dotted lines, corresponding to the isotropic-nematic and nematic-rhombatic transitions, respectively. Lastly, we added an additional vertical green dotted line, which corresponds to a sudden rise in all order parameters around $\eta \approx 0.91$. This observation may signify the formation of a solid phase.

We note that the melting process occurs in three distinct steps, as the order in the particles' orientation and the bond order emerges at different densities. This process is reminiscent of the melting of hard superdisks~\cite{gurin2020three}. However, there is a difference, which can be attributed to the more pronounced anisotropy of the rhombus shape compared to that of superdisks. In the latter case, as the density increases, the complete orientational symmetry of the isotropic phase is initially broken by the quasi-long-range bond order, forming a hexatic phase. Subsequently, the particles' orientational symmetry is disrupted, resulting in a rhombatic phase. For rhombi, the order of these symmetry breakings is reversed. At lower densities, the particles' orientational symmetry is broken, leading to a nematic phase, and at higher densities, the bond order symmetry is disrupted, resulting in a rhombatic phase before reaching the solid state.

\begin{figure*}[t!]
\centering
\includegraphics[width=0.9\linewidth]{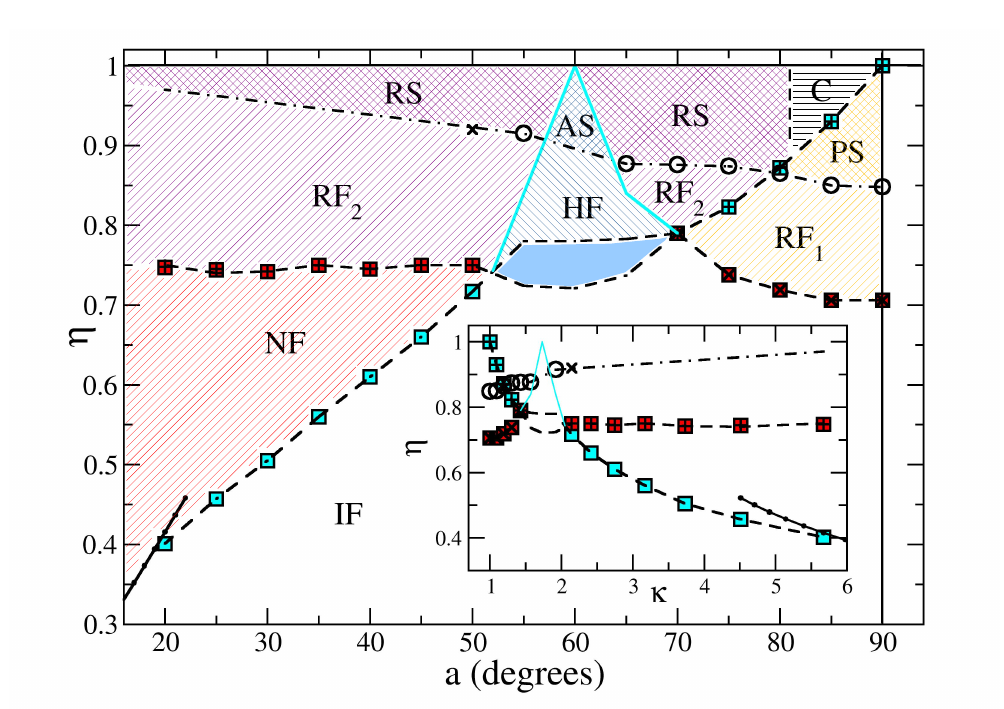}
\caption{Phase diagram of hard rhombi. IF - Isotropic fluid, NF - Nematic fluid, RF$_1$ - Rhombatic fluid with low $P_2$, RF$_2$ - Rhombatic fluid with high $P_2$, HF - Hexatic fluid, C - Columnar phase, RS - Rhombic solid, AS - Aperiodic solid, and PS -Plastic (rotator) solid. Cyan and red squares represent sharp increases in $P_2$ and $P_4$ respectively, crosses and plus signs correspond to sharp increases in $\Psi_4$ and $\Psi_6$, dashed lines indicate $\chi$ peaks, and circles mark solid-fluid boundaries, determined from quasi-long range correlations with $N=5000$. Regions with 45-degree hatching represent non-isotropic fluid regions, while regions with intersecting hatching correspond to solid phases. The horizontally hatched area denotes the columnar phase. The cyan solid area is the coexistence between isotropic and hexatic fluids. The bullets along with the solid black line correspond to the function $3/4\pi \tan(a/2)$, representing an extrapolation of the Onsager limit for hard needles. The inset shows the same data in the $\eta$-$\kappa$ plane, where $\kappa=1/\tan(a/2)$ is the particle anisotropy.  }
\label{fig:phase-diagram}
\end{figure*}

As previously mentioned, the black and red dashed curves introduced in panels b) and c) of Fig.~\ref{fig:EOS-a40} correspond to $a=45^\circ$ and $a=35^\circ$, respectively. While the $\eta \approx 0.75$ peak in the $\chi$ curve remains unchanged, the first peak exhibits a substantial shift to lower $\eta$ as $a$ decreases. This observation confirms our initial expectation that as the particles become needle-shaped, the extent of their nematic region expands in $\eta$ approximately proportional to $a$ (thus, in $\rho$ is approximately constant). More precisely, the constant value of $\rho_{IN}L^2$ implies that $\eta_{IN}=\rho_{IN} L^2/2 \tan(a/2) \approx \frac{1}{4} \rho_{IN} L^2 a$ as $a \rightarrow 0$, where $L=2\sigma \cos(a/2)$ represents the length of the needle-like shape and $\rho_{IN}$ denotes the corresponding number density at which the isotropic nematic transition occurs at the needle limit. By adopting the estimation from the Onsager theory~\cite{kayser1978}, $\rho_{IN}=3\pi/2 \approx 4.71$, yielding $\eta_{IN} \approx 0.41$ at $a=20^\circ$, consistent with our simulation results. However, it's important to note that our data suggests a slower decay of $\eta$ with decreasing $a$ compared to $3\pi/4 \tan(a/2)$ (this function is depicted in Fig.~\ref{fig:phase-diagram} as bullets), at least around $a\approx 20^\circ$. Thus, we think our results should surpass the Onsager value of $3\pi/4$ for the needle limit, aligning with simulations~\cite{Frenkel-Eppenga_PRA_1985}. Interestingly, the transition density from nematic to rhombatic phases, which appears to be minimally impacted by the particles' shape in $\eta$, diverges in $\rho$ approximately as $1/\sin(a)$.

\begin{figure*}[t!]
\centering
\includegraphics[width=0.9\linewidth]{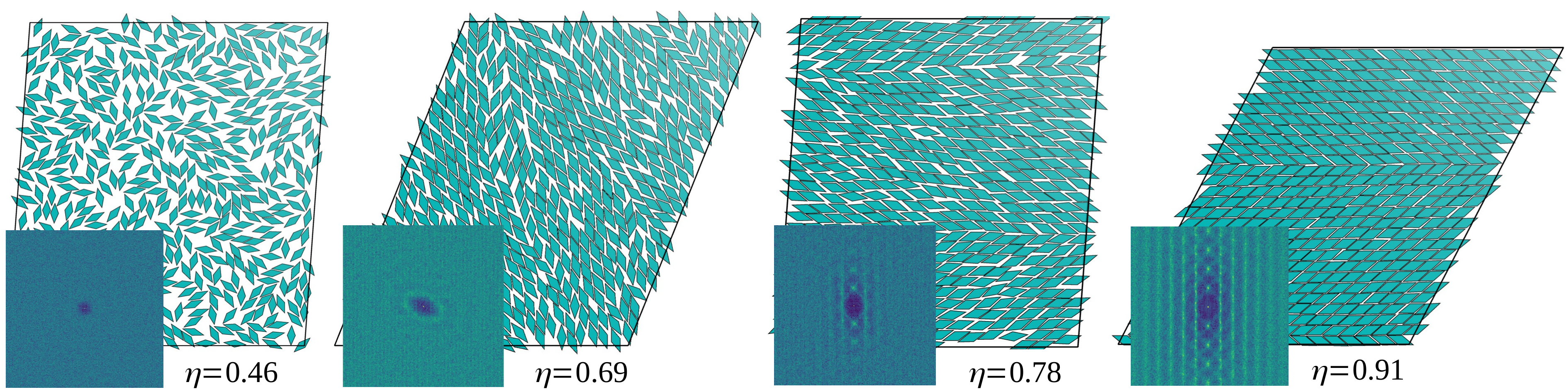}
\caption{Snapshots and their corresponding SSFs (computed by using freud~\cite{ramasubramani2020freud}) for rhombi with $a=40^\circ$ at increasing packing fraction. The first (leftmost) panel shows an isotropic fluid, the second a nematic phase, the third a poly-rhombatic phase, and the fourth a poly-rhombic phase.  }
\label{fig:snapshots-a40}
\end{figure*}

Fig.~\ref{fig:snapshots-a40} presents snapshots and SSFs of rhombi with $a=40^\circ$ at varying densities, progressing from left to right. The leftmost panel corresponds to a typical isotropic configuration, while the second panel showcases a nematic phase. The third snapshot exhibits regions where particles are oriented in both left and right directions. If one of these regions were to occupy space entirely, the snapshot would depict a rhombatic fluid. A similar pattern is discernible in the rightmost panel, where two distinct domains corresponding to different rhombic tilings are evident. It's important to note that in these $N=500$ systems, we've also encountered numerous snapshots featuring a single domain representing a true rhombatic or rhombic configuration, which we believe corresponds to equilibrium.

It's noteworthy that establishing boundaries between these domains appears to incur minimal entropy cost. Something similar occurs when creating an ABAB 3D stacking of triangular planes (an FCC arrangement), an ABCABC stacking (an HCP configuration), or any disordered ABABC stacking. While the ABAB 3D configuration may possess marginally lower entropy, the disparity is almost negligible~\cite{kratky1981stability,sanchez2021fcc}. As a result, we could anticipate the emergence of poly-rhombic and poly-rhombatic phases for large system sizes close to equilibrium. In fact, when initiating from loosely random conditions, we consistently observe these types of configurations in our $N=5000$ systems.

\begin{figure}[t!]
\centering
\includegraphics[width=0.9\linewidth]{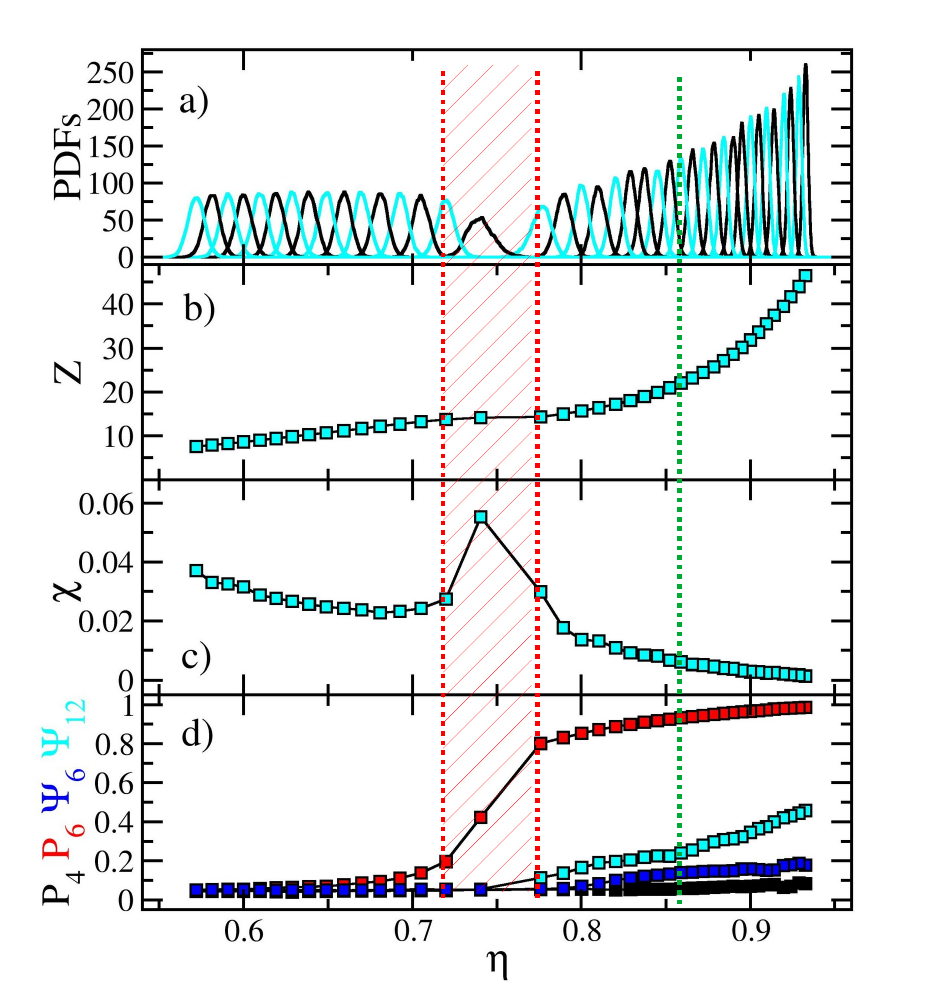}
\caption{a) Probability density functions, b) compressibility factor, c) dimensionless isothermal compressibility, and d) order parameters as a function of the packing fraction for rhombi with a minor angle $a=60^\circ$. In panel d), black and red symbols depict orientational order parameters $P_4$ and $P_6$, respectively, while blue and cyan symbols correspond to bond-orientational order parameters $\Psi_6$ and $\Psi_{12}$, respectively. The red vertical dotted lines highlight the region where a first-order transition takes place. The green vertical dotted line indicates the point where we believe the solid melts. }
\label{fig:EOS-a60}
\end{figure}

Let's now examine the unique scenario of rhombi with a minor angle of $a=60^\circ$, as previously studied by Whitelam et al.~\cite{Whitelam2012}. For this case, the associated PDFs, $Z$, $\chi$, and order parameters are displayed in Fig.~\ref{fig:EOS-a60}. Note that in panel d), we show $P_6$ instead of $P_2$ and $\Psi_{12}$ instead of $\Psi_4$ because we expect a six-fold symmetry for the particle orientation and found relatively large $\Psi_{12}$ values at high densities. We have also measured $\Psi_{4}$, $\Psi_{8}$, and $\Psi_{10}$ in this case, but we only found $\Psi_{6}$ and $\Psi_{12}$ to significantly grow with increasing density.

\begin{figure*}[t!]
\centering
\includegraphics[width=0.9\linewidth]{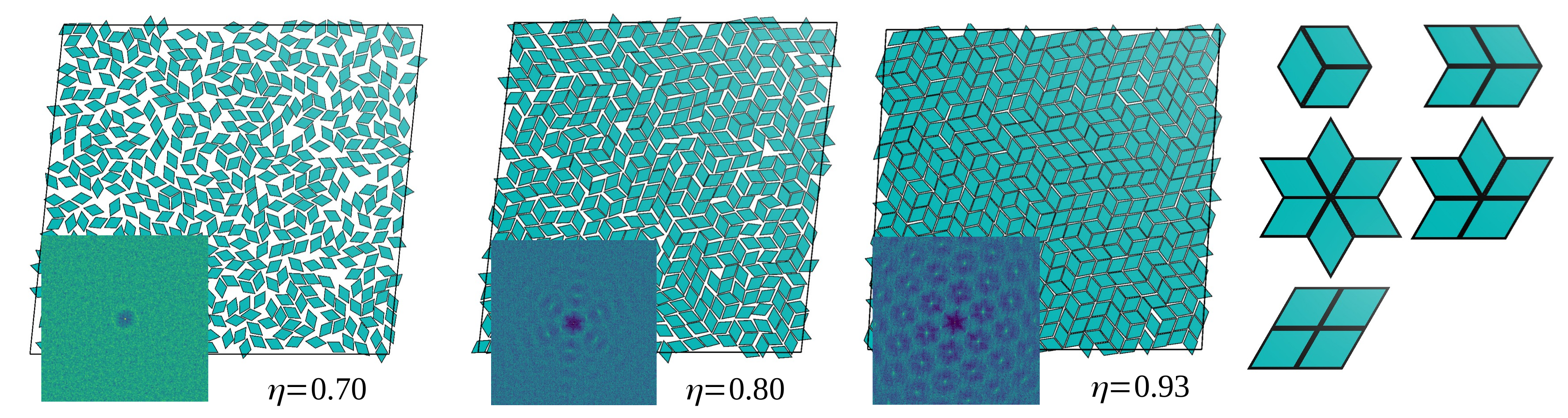}
\caption{Snapshots and their corresponding SSFs (computed by using freud~\cite{ramasubramani2020freud}) for rhombi with $a=60^\circ$ at increasing packing fraction. The first (leftmost) panel shows an isotropic fluid, the second a hexatic fluid, and the third is an aperiodic solid. The five different clusters depict all the possible local arrangements (up to rotations by 60 degrees) around a shared corner. These clusters naturally appear in the high-density packing \cite{Stannard2012}.   }
\label{fig:snapshots-a60}
\end{figure*}

The compressibility factor $Z$ exhibits a distinct plateau, while $\chi$ displays a prominent peak at approximately $\eta = 0.74$, coinciding with the point where the PDFs deviate from their Gaussian distribution. These characteristics serve as indicators of a first-order phase transition. In addition, we observe that $P_6$ strongly grows at $\eta = 0.74$ while the other order parameters remain modest and undergo minimal changes as the system undergoes this transition. Indeed, $P_6$ emerges near $\eta = 0.7$, consistent with the value reported elsewhere for the same system~\cite{Whitelam2012}.

To uncover the phase into which the isotropic fluid is transitioning, an examination of the snapshots and SSFs proves illuminating. These images are illustrated in Fig.~\ref{fig:snapshots-a60}, unveiling a six-fold orientationally ordered fluid state. Note that this fluid is completely analogous to the nematic phase, as it is orientationally ordered, has no bond nor positional correlations, but features six-fold instead of two-fold symmetry. Hence, the corresponding phase should be called hexatic~\cite{Whitelam2012}, although it conflicts with the well-accepted hexatic phase associated with disks, where neighboring particles bond showing a six-fold symmetry. Something similar happens with the tetratic phase associated with squares where bond and orientational four-fold symmetry appear.

Subsequently, the hexatic fluid (let us call it this way) solidifies, with a slow increasing of $\Psi_6$ and $\Psi_{12}$. Upon compression, this solid can completely and aperiodically fill space, hindering the growth of quasi-long-range positional correlations. In addition, it is clear from the images for $\eta=0.93$ that the lack of translational symmetry and a the discreteness of the SSF image are evident, both necessary conditions to classify the phase as a quasicrystal~\cite{levine1986quasicrystals,toudic2008hidden,janssen2018aperiodic,sosa2023structural}. It also appears that there are motifs that are repeated aperiodically through space (see the right panel of Fig.\ref{fig:snapshots-a60}). However, there are other required conditions an aperiodic tiling must fulfill to be considered a quasicrystal\cite{levine1986quasicrystals,toudic2008hidden,janssen2018aperiodic,sosa2023structural}. On the other hand, one should note that by randomly joining the equilateral triangles of a triangular tessellation to produce rhombi, as done with disks to form dimers~\cite{Wojciechowski1991} or with squares to form rectangles~\cite{Donev2006}, one obtains a type of aperiodic random tiling. Thus, we believe we should call this phase an aperiodic solid (or non-periodic solid). Finally, note that the centers of the triangles are distributed on a Kagome lattice, like the centers of the dimers formed by joining disks. Thus, the disk and rhombic tessellations yielded at infinite pressure are isomorphic, meaning that their residual entropies are the same. As pointed out by Wojciechowski et al., the residual entropy per particle of this solid at infinite pressure is around 0.857$k_B$~\cite{Wojciechowski1991,fowler1937attempt,phares1986thermodynamics}.

The observation that $\Psi_{12}$ surpasses $\Psi_6$ can be elucidated by considering the number of neighbors within an equilateral triangular tessellation. When we take into account both the first and second neighbors instead of solely the first ones, $\Psi_{6}$ approaches zero while $\Psi_{12}$ equals one. This phenomenon arises because six of the twelve neighbors are rotated by an angle of $\pi/6$ relative to the other six. To some extent, a similar effect is observed in the $a=60^\circ$ rhombi system, providing an explanation for the observation of $\Psi_{12} > \Psi_6$.

We speculate that the solid might transform into the hexatic fluid phase around $\eta \approx 0.86$, but we currently lack proof of this conjecture. The absence of discernible positional correlations impedes the MC simulations to estimate this phase transition boundary, leading us to believe that only dynamic simulations could provide an alternative mean to precisely determine this point. In addition, we may further speculate that this transition is not thermodynamic but dynamic, as there is not any discernible structural change, but the smooth growing of $\Psi_{6}$ and $\Psi_{12}$ and the gain in definition of the SSF. This speculation is also supported as neither the PDFs, nor the $Z$, or $\chi$ functions display any kind of signature.

In the rightmost panel of Fig.~\ref{fig:snapshots-a60}, we showcase various configurations of $a=60^\circ$ rhombi, encompassing a 2D cube-like arrangement, a star-like pattern, a magnified rhombus four times the particle size, an arrow-like layout, and a hybrid design fusing elements of the star and the large rhombus. These figures align perfectly with each other, maintaining their arrangement even after undergoing $60^\circ$ rotations for all clusters.  These intricate patterns spontaneously manifest within the snapshot at $\eta \approx 0.93$. While the system's periodic boundary conditions enforce periodicity, we can anticipate the emergence of aperiodic arrangements in the thermodynamic limit. In fact, within the supplementary material (SM), we provide a snapshot and its corresponding SSF obtained from starting with random loose configurations showcasing aperiodic patterns within the confines of the $N=5000$ system cell (Fig. S1).

Finally, we can compare the results obtained from $a=60^\circ$ rhombi with those for equilateral triangles, as a pair of the latter can combine to form a single rhombus. As for squares, the observed transition of triangles from the isotropic fluid to the $i$-atic phase is continuous, following the KTHNY scenario~\cite{Anderson2017}. Moreover, equilateral triangles exhibit chiral phases at high densities~\cite{zhao2012,gantapara2015}. Hence, it is evident that the behavior of equilateral triangles significantly differs from that of $a=60^\circ$ rhombi, akin to the distinction observed with disks and homonuclear dimers~\cite{Wojciechowski1991}, as well as squares and the rectangles formed by their combination~\cite{Donev2006}.

For systems with $55^\circ \leq a \leq 65^\circ$, configurations resembling the aperiodic solid spontaneously emerge as $\eta$ values become sufficiently large. In these cases, $P_6$ jumps and $\Psi_{12}$ starts growing with increasing $\eta$. Based on the shape of the PDFs, the extension of the $Z$ plateaus, and the magnitude of the $\chi$ peaks, we posit that the systems with these $a$ values undergo an isotropic-hexatic first-order transition. As $\eta$ continues to rise, we expect the hexatic fluid to solidify into an aperiodic solid (or become a glass). However, the exact location of this transition remains to be determined. Finally, as $\eta$ approaches unity, and considering that the sole aperiodic tiling capable of completely filling space occurs at $a=60^\circ$, we speculate the existence of a solid-solid transition leading to the formation of a space-filling rhombic solid.

Within the supplementary material (SM), we present the PDFs, $Z$, $\chi$, and order parameters for the cases of $a=55^\circ$ and $65^\circ$, starting from loose random configurations (see Fig. S2 and Fig. S3). These plots also encompass the $Z$, $\chi$, and order parameters derived from initiating with compact rhombic solids, highlighting the establishment of glassy states. This observation implies that the system's history influences the eventual steady state, indicating the absence of equilibrium for $\eta>0.85$. It's noteworthy that the transition from an aperiodic configuration to a periodic rhombic solid is significantly hindered, explaining the failure of the REMC technique to attain equilibrium in this scenario. In the SM, we also show high-density snapshots and their corresponding SFFs as obtained by starting from random configurations from $a=50^\circ$ to $a=70^\circ$ in steps of $5^\circ$ (Fig. S4). Note that $a=50^\circ$ and $a=70^\circ$ yield rhombic solids (for these cases, the EOS does not depend on the initial configurations for all $\eta$).

By compiling all the gathered information, we construct what we believe to be the equilibrium phase diagram of hard rhombi. The comprehensive overview is presented in Fig.~\ref{fig:phase-diagram}, where cyan and red squares indicate sharp increases in $P_2$ and $P_4$ respectively, crosses and plus signs correspond to pronounced rises in $\Psi_4$ and $\Psi_6$, dashed lines represent $\chi$ peaks, and circles denote solid-fluid boundaries, as determined from quasi-long range correlations with $N=5000$. The fluid regions are indicated by single hatching for non-isotropic fluids, while double hatching is applied to depict solid phases. The cyan-shaded region corresponds to an isotropic-hexatic coexistence. In alignment with Fig.~\ref{fig:snapshots-a85}, a relatively small region is outlined in the upper right corner, suggesting a potential columnar (or tilted smectic, taking the mayor diagonal as the particle main axis) phase. However, we acknowledge that further research is necessary to confirm the existence of this phase.

At $a=90^\circ$, the phase diagram aligns with the behavior of hard squares~\cite{Anderson2017}. However, as $a$ departs from $90^\circ$ towards slightly lower values, the square solid transforms into a plastic solid, $PS$, and the tetratic fluid into a rhombatic fluid, $RF_1$. To differentiate it from another rhombatic fluid characterized by a higher $P_2$, we label this other phase as $RF_2$, the $RF_1$ phase consists of particles that can easily rotate by $90^\circ$. This kind of rotation is significantly hindered in the $RF_2$ phase. Note that there is a marked $\chi$ peak associated to the $RF_1$-$RF_2$ fluid-fluid transition (see Fig. S8 of the SM). As $a$ slightly departs from $90^\circ$, $\eta$ can be increased close to unity with a low $P_2$, enabling particles to perform $90^\circ$ rotations. For clarity, we refer to this phase as the "plastic" phase, although the term "rotator" might be more appropriate (using another "RS" abbreviation in the phase diagram could lead to confusion). It's important to note that while the rotator phase permits $90^\circ$ rotations (feasible in MC simulations), continuous rotations are likely somewhat restricted, in contrast to the "fully" rotator phases observed in low-anisotropy ellipses~\cite{Cuesta1990,Donev05b,Bautista14}. Upon increasing $\eta$ towards 1, particles within this phase align, and $P_2$ increases, resulting in the formation of either a rhombic solid or a columnar phase. Both of these structures possess the capability to completely occupy space.

As $a$ decreases, the isotropic fluid becomes favored over $RF_1$. Furthermore, the boundary of $RF_2$ shifts towards lower $\eta$, also contributing to the reduction of the $RF_1$ region, which ultimately disappears at $\eta$ around $a=70^\circ$. Beyond this point, for even lower $a$ values, $RF_2$ stands as the sole prevailing rhombatic phase. Setting aside the emergence of aperiodic phases at around $a=60^\circ$, the decrease in $a$ promotes the initial appearance and subsequent expansion of the nematic fluid region. This region grows as $a$ decreases, ultimately encompassing the entire range of $\eta$ as $a$ approaches zero, aligning with the anticipated needle-like behavior~\cite{Onsager_AnnNYAcadSci_1949,Frenkel-Eppenga_PRA_1985,tarjus1991new,Vink2009,Gurin_PRE_2011}.

It's important to note that simulations employing $N=5000$ particles and starting from rhombic solid configurations fail to preserve quasi-long range positional correlations even for $\eta=0.91$ with $a=30^\circ$ (see Fig. S5 in the SM), in contrast to the case of $\eta=0.874$ with $a=75^\circ$ (see Fig. S6 in the SM). Consequently, we conjecture that the $RS$ phase diminishes in its $\eta$ range as $a$ decreases, ultimately vanishing at $a=0^\circ$. Another noteworthy observation is that both the $Z$ plateau and the $\chi$ peak corresponding to the nematic-rhombatic fluid-fluid transition decrease their size as $a$ decreases (see the SM for the PDFs, $Z$, $\chi$, and global order parameters for $a=20^\circ$). Although we still observe a small $\chi$ peak for $a=20^\circ$, its diminishing trend suggests its disappearance as $a$ approaches zero (see Fig. S7). Hence, the limit as $a$ tends to zero remains in complete accordance with the expected needle-like behavior.

The inset of Fig.\ref{fig:phase-diagram} presents the same data as the main panel but in the $\eta$-$\kappa$ plane, where $\kappa=1/\tan(a/2)$ represents the particle anisotropy. This inset allows for a direct comparison with the phase diagrams of hard spherocylinders~\cite{Bates2000} and ellipses~\cite{Bautista14}. Setting aside the aperiodic solid and the hexatic fluid observed for rhombi, several common features emerge. In all cases, close to $\kappa=1$, there is a high-density plastic (rotator) solid phase, and no nematic phase is present. Additionally, there is always a minimum $\kappa$ value at which the nematic phase emerges and expands its $\eta$ domain. For rhombi, we estimate this value to be approximately 2.1, while for ellipses, it's around 2.4~\cite{Bautista14}, and for spherocylinders, it falls in the range of 6.0-9.0~\cite{Bates2000}.

Around $a=60^\circ$, the phase diagram becomes disrupted due to the emergence of an aperiodic solid phase resembling the one found for disk-dimers~\cite{Wojciechowski1991} and rectangles with $\kappa=2$~\cite{Donev2006}. This aperiodic solid subsequently undergoes a soft transition to a hexatic fluid phase before finally yielding the isotropic fluid through a first-order transition. This melting behavior shares similarities with that observed for disks~\cite{Bernard_PRL_2011}, particularly in terms of the order of the subsequent transitions and the appearance of a hexatic phase (although in terms of particle orientation instead of neighbors), yet it starkly contrasts as it involves the melting of an aperiodic structure. This is a remarkable distinction, given that there is six-fold symmetry in particle orientation and twelve-fold symmetry for bonds to be broken.

The aperiodic pattern spontaneously emerges within the confines of our system boundaries, even when initializing with a rhombic solid configuration. The substantial entropy gain associated with this structure, relative to its periodic competitors (such as rhombic and tumbling blocks), likely accounts for its spontaneous formation, even for cases like $a=55^\circ$ and $a=65^\circ$, where slight mismatches arise avoiding efficiently filling space.

We would like to emphasize that there are regions where we have observed a three-step melting scenario, specifically along the $RS$-$RF_2$-nematic and the $RS$-$RF_2$-$RF_1$ pathways with decreasing density. These pathways resemble the melting behavior observed in superdisks for shapes that are intermediate between disks and squares~\cite{gurin2020three}. In all cases, the $RS$ phase initially transitions into an rhombatic fluid, which inherits its local symmetry. However, the subsequent melting processes differ. For relatively low $a$ values, the local rhombic symmetry is lost, leaving only the nematic director preserved, whereas for large $a$ values, the rhombic symmetry is retained, but the particles easily begin to exhibit 90-degree turns while sharply decreasing the $P_2$ parameter. Recalling the behavior of superdisks, the rhombatic fluid leads to a hexatic phase, where only the six-fold bond order maintains its quasi-long-range correlations. Hence, it appears that three-step melting scenarios are not so uncommon for systems of hard particles in two dimensions.

\section{Conclusions}

We have investigated the phase boundaries of bulk 2D rhombi through replica exchange Monte Carlo simulations. To achieve this, we primarily conducted simulations using relatively small system cells with $N=500$ particles. Additionally, we carried out some supplementary simulations with $N=5000$ particles to estimate the fluid-solid boundaries. The resulting phase diagram is presented in the $\eta$-$a$ and $\eta$-$\kappa$ planes, where $\eta$ represents the packing fraction density, $a$ signifies the minor angle of the rhombus, and $\kappa$ is the particle anisotropy. Our analysis is based on various data sources, including order parameters, the compressibility factor, $Z$, dimensionless isothermal compressibility, $\chi$, and observations from snapshots and SFFs.

The phase diagram is depicted in Fig.~\ref{fig:phase-diagram}, illustrating an isotropic fluid phase governing the low-density region. This phase keeps its $\eta$ range as $a$ increases up to approximately $a=50^\circ$. At lower values of $a$, the isotropic fluid undergoes a transition as $\eta$ increases, first to a nematic phase and then to a rhombatic fluid, eventually solidifying. Around $a=60^\circ$, a first-order transition occurs in the isotropic fluid, leading to a six-fold orientationally ordered fluid, a hexatic phase, which subsequently smoothly transforms into an aperiodic solid.

Beyond $a=70^\circ$, the isotropic fluid transforms into a rhombatic fluid with initially low $P_2$ (two-fold orientational order parameter), which then increases, culminating in the formation of the rhombic solid phase at higher densities. For $a>80^\circ$, the rhombatic fluid with low $P_2$ undergoes solidification before experiencing an increase in $P_2$, allowing particles to rotate freely in 90-degree steps, a behavior akin to a rotator or plastic solid. At even higher densities, $a>80^\circ$ rhombi exhibit a transition to a (possible) columnar phase.

In summary, the bulk hard rhombi system has a rich phase diagram composed of various solid and fluid phases.

\section{Data availability}

The raw/processed data required to reproduce these findings cannot be shared at this time due to technical or time limitations.

\section{Acknowledgements}

The authors express their gratitude for the financial support provided by the CONAHCyT project A1-S-9197. We also wish to extend our appreciation for the insightful discussions with Atahualpa S. Kraemer on the subject of quasicrystals, as well as with Szabolcs Varga. P.G. gratefully acknowledges the financial support of the National Research, Development, and Innovation Office - Grant No. NKFIH K137720 and TKP2021-NKTA-21.


%

\end{document}